\documentclass[11pt,twoside]{article}
\usepackage{graphicx}
\usepackage{amsmath}
\usepackage{amssymb}
\usepackage{cite}

\begin{document}
\newcommand{\pst}{\hspace*{1.5em}}

\newcommand{\rigmark}{\em Journal of Russian Laser Research}

\newcommand{\be}{\begin{equation}}
\newcommand{\ee}{\end{equation}}
\newcommand{\bm}{\boldmath}
\newcommand{\ds}{\displaystyle}
\newcommand{\bea}{\begin{eqnarray}}
\newcommand{\eea}{\end{eqnarray}}
\newcommand{\ba}{\begin{array}}
\newcommand{\ea}{\end{array}}
\newcommand{\arcsinh}{\mathop{\rm arcsinh}\nolimits}
\newcommand{\arctanh}{\mathop{\rm arctanh}\nolimits}
\newcommand{\bc}{\begin{center}}
\newcommand{\ec}{\end{center}}
\newcommand{\ket}[1]{{\left| #1 \right>}}
\newcommand{\bra}[1]{{\left< #1 \right|}}
\newcommand{\aver}[1]{{\left< #1 \right>}}

\thispagestyle{plain}

\label{sh}


\begin{center} {\Large \bf
\begin{tabular}{c}
GEOMETRIC OPTICS WITH ATOMIC BEAMS 
\\[-1mm]
SCATTERED BY A DETUNED STANDING LASER WAVE
\end{tabular}
 } \end{center}

\bigskip

\bigskip

\begin{center} {\bf S.V.~Prants, V.O. Vitkovsky, L.E. Konkov}
\end{center}

\medskip

\begin{center}
{\it
Laboratory of Nonlinear Dynamical Systems,\\
Pacific Oceanological Institute of the Russian Academy of Sciences,\\
690041 Vladivostok, Russia, URL: dynalab.poi.dvo.ru}

\smallskip

$^*$Corresponding author e-mail:~~~prants@poi.dvo.ru
\\
\end{center}
\begin{abstract}\noindent
We report on theoretical and numerical study of propagation of atomic beams 
crossing a detuned standing-wave laser beam in the geometric optics limit. 
The interplay between external and internal atomic degrees of freedom is used to 
manipulate the atomic motion along the optical axis by light. By adjusting the atom-laser detuning, 
we demonstrate how to focus, split and scatter atomic beams in a real 
experiment. The novel effect of chaotic scattering of atoms at a regular 
near-resonant standing wave is found numerically and explained qualitatively. 
Some applications of the effects found are discussed. 
\end{abstract}

\medskip

\noindent{\bf Keywords:}
atomic scattering, standing wave, optical nanolithography

\section{Introduction}
\pst
Manipulation of atoms by light becomes possible due to 
the dipole forces which are well described by the semiclassical model 
with quantum description of internal atomic transitions induced by a 
near resonant laser field and classical description of their center-of-mass motion 
\cite{Kazantsev}. For the first time, the ideas to trap and channel 
cold atoms with the help of standing laser waves (SLW) have been proposed by 
V.~Letokhov and his co-workers \cite{L68,BL87,Letokhov}. The ability 
of a SLW to deflect, 
channel and split atomic beams \cite{Arimondo,Mlynek}
has been used for a variety of applications including atom microscopy, 
interferometry, isotope separation and optical lithography   
\cite{Mlynek,Sleator,Timp,McClelland}. Lasers can be used to manipulate atomic trajectories to create 
atomic analogues of such familiar optical phenomena as 
focusing of light, beam splitting and light scattering. 
It is remarkable that now we are able to reverse the roles of light and 
matter from their familiar roles. 
The semiclassical description, used in this paper, is similar to the 
geometric optics limit in conventional optics. Atomic trajectories play 
the role of light rays with the SLW being a light mask. 

In the present paper we intend to demonstrate theoretically and numerically 
that adjusting  in an experiment only one parameter, 
the detuning between the frequencies of a working atomic transition and 
the SLW, one can explore a variety of the regimes of the atom-laser 
interaction to focus, split and scatter atomic beams. 
Near the atom-field resonance, where  
the interaction between the internal and external atomic degrees of freedom is 
intense, there is a possibility to create conditions for chaotic  
scattering of atoms \cite{JETPL01,PRE02,PRA01} without any additional efforts 
like a SLW modulation. It becomes possible due to the peculiarities 
of the dipole force in a near-resonant optical lattice 
\cite{JETP03,JRLR06,PRA07,PRA08}.

\section{Focusing, splitting, bunching and scattering of atomic beams}
\subsection{Equations of motion}
\pst

A beam of two-level atoms in the $z$ direction crosses  
a SLW laser field with optical axis in the $x$ direction. 
The laser beam has the Gaussian profile
$\exp[-(z-z_0)^2/r^2]$ with $r$ being the $e^{-2}$ radius at the laser beam waist.
The characteristic length of the atom-field interaction is supposed to be 
$\pm 1.5r$ because the light intensity at $z=\pm 1.5r$ is two orders of 
magnitude smaller than the peak value. The longitudinal velocity of atoms, 
$v_z$,  
is much larger than their transversal velocity $v_x$ and is supposed to be 
constant. Thus, the spatial laser profile may be replaced by the temporal one. 
The Hamiltonian of a two-level atom in the one-dimensional SLW  
can be written in the frame rotating with the angular laser frequency $\omega_f$ 
as follows:
\begin{equation}
\hat H=\frac{P^2}{2m_a}+\frac{\hbar}{2}(\omega_a-\omega_f)\hat\sigma_z-\\
\hbar \Omega_0\exp[-(t-\frac{3}{2} \sigma_t)^2/\sigma^2_t]
\left(\hat\sigma_-+\hat\sigma_+\right)\cos{k_f X_a} 
-\frac{i\hbar\Gamma}{2}\hat\sigma_+\hat\sigma_-,
\label{Ham}
\end{equation}
where $\hat\sigma_{\pm, z}$ are the Pauli operators for the internal atomic degrees
of freedom, $X_a$ and $P$ are the classical atomic position and momentum, 
$\Gamma$, $\omega_a$, and $\Omega_0$ are the decay rate, the atomic 
transition and maximal Rabi frequencies, respectively.  
The simple wave function for the electronic degree of freedom is
$\ket{\Psi(t)}=a(t)\ket{2}+b(t)\ket{1}$, where $a \equiv A+i\alpha$ and $b\equiv B+i\beta$ are the complex-valued probability 
amplitudes to find the 
atom in the excited, $\ket{2}$, and ground, $\ket{1}$, states, respectively.

In the semiclassical approximation, atom with quantized internal dynamics is  
treated as a point-like particle to be described by the Hamilton--Schr\"odinger 
equations of motion written for the real and imaginary parts of the probability 
amplitudes 
\begin{equation}
\begin{gathered}
\dot x =\omega_r p, \, \dot p =- 2\exp[-(\tau-\frac{3}{2} 
\sigma_{\tau})^2/\sigma^2_{\tau}](AB+\alpha \beta)\sin x, 
\\ 
\dot A = \frac12 (\omega_rp^2 - \Delta)\alpha - \frac12 \gamma A 
- \exp[-(\tau-\frac{3}{2} \sigma_{\tau})^2/\sigma^2_{\tau}]\beta\cos x,
\\
\dot \alpha = -\frac12 (\omega_rp^2 - \Delta)A - \frac12 \gamma 
\alpha + \exp[-(\tau-\frac{3}{2} \sigma_{\tau})^2/\sigma^2_{\tau}]B\cos x, 
\\
\dot B = \frac12 (\omega_rp^2 + \Delta)\beta -  
\exp[-(\tau-\frac{3}{2} \sigma_{\tau})^2/\sigma^2_{\tau}]\alpha \cos x,
\\
\dot \beta = -\frac12 (\omega_rp^2 + \Delta)B+ 
\exp[-(\tau-\frac{3}{2} \sigma_{\tau})^2/\sigma^2_{\tau}]A\cos x, 
\end{gathered}
\label{mainsys}
\end{equation}
where $x\equiv k_f X_a$ and $p\equiv P/\hbar k_f$ are scaled atomic center-of-mass position 
and transversal momentum, respectively and dot denotes differentiation with respect 
to the dimensionless time $\tau\equiv \Omega_0 t$. The recoil frequency, 
$\omega_r\equiv\hbar k_f^2/m_a\Omega_0\ll 1$,
the atom-laser detuning, $\Delta\equiv(\omega_f-\omega_a)/\Omega_0$, 
the decay rate $\gamma=\Gamma/\Omega_0$, and the characteristic interaction time, 
$\sigma_{\tau}\equiv r\Omega_0/v_z$, are the control parameters. 

Let us introduce
instead of the complex-valued probability amplitudes $a$ and $b$ the
following real-valued variables:
\begin{equation}
u\equiv 2\operatorname{Re}\left(ab^*\right),\quad
v\equiv -2\operatorname{Im}\left(ab^*\right), \quad
z\equiv \left|a\right|^2-\left|b\right|^2,
\label{uvz_def}
\end{equation}
where $u$ and $v$ are synchronized and quadrature
components of the atomic electric dipole moment, respectively, and $z$ is
the atomic population inversion.
In the absence of any losses ($\gamma=0$), Eqs.~(\ref{mainsys}) 
can be cast in the form 
\begin{equation}
\begin{gathered}
\dot x=\omega_r p,\quad \dot p=- u\exp[-(\tau-\frac{3}{2} 
\sigma_{\tau})^2/\sigma^2_{\tau}]\sin x, \quad \dot u=\Delta v,\\
\dot v=-\Delta u+2\exp[-(\tau-\frac{3}{2} 
\sigma_{\tau})^2/\sigma^2_{\tau}] z\cos x, \quad
\dot z=-2\exp[-(\tau-\frac{3}{2} 
\sigma_{\tau})^2/\sigma^2_{\tau}] v\cos x.
\end{gathered}
\label{Hamsys}
\end{equation}
The system (\ref{Hamsys}) has two integrals of motion, namely, the total energy
\begin{equation}
H\equiv\frac{\omega_r}{2}p^2-u\cos x-\frac{\Delta}{2}z,
\label{H}
\end{equation}
and the length of the Bloch vector, $u^2+v^2+z^2=1$,
whose conservation follows immediately from
Eqs.~(\ref{uvz_def}).

Equations (\ref{Hamsys}) constitute a nonlinear Hamiltonian
autonomous system with two and half degrees of freedom which, owing 
to the two integrals of motion, move on a three-dimensional
hypersurface with a given energy value $H$. In general, motion in 
a three-dimensional phase space in characterized by a positive
Lyapunov exponent $\lambda$, a negative exponent equal in magnitude to the positive one,
and zero exponent \cite{KP97}.
The maximal Lyapunov exponent characterizes the mean rate of the
exponential divergence of initially close trajectories and serves as 
a quantitative measure of dynamical chaos in the system.
The values of the maximal Lyapunov exponent in dependence on
the detuning, the recoil frequency and the initial atomic momentum
have been computed in Refs.~\cite{JRLR06,PRA07}.

There are different regimes of the center-of-mass motion 
along the SLW optical axis \cite{PRA01,PRA07}. In dependence on the initial conditions 
and the values of the control parameters, atoms may oscillate in a regular or a 
chaotic way in wells of the optical potential or move ballistically 
over its hills with regular or chaotic variations of their velocity. 
Chaotic motion with a positive value of the maximal Lyapunov exponent  
becomes possible in a narrow range of the detuning values, 
$0< |\Delta| <1$~\cite{PRA07}. 
At $\Delta=0$, the synchronized electric-dipole component, 
$u$, becomes a constant. That implies the additional integral 
of motion in the Hamiltonian version (\ref{Hamsys}) of Eqs.~(\ref{mainsys}) 
and the regular motion with zero maximal Lyapunov exponent. 
Far off the resonance, at $|\Delta| >1$, the motion is regular both in the trapping 
and ballistic modes. 

It is remarkable that there is a specific type of motion, chaotic walking in 
a deterministic optical potential, when atoms can change the direction 
of motion alternating between flying through 
the SLW and being trapped in its potential wells. 
We would like to stress that the local instability produces chaotic center-of-mass 
motion in a rigid SLW without any modulation of its parameters. 
Chaotic walking occurs due to the specific 
behavior of the Bloch-vector component  
of a moving atom $u$ whose shallow oscillations between the SLW nodes are interrupted by 
sudden jumps with different amplitudes while atom crosses each node of the SLW 
\cite{PRA07}. It looks like a random like shots happened in a fully 
deterministic environment. It follows from the second 
equation in the set ~(\ref{Hamsys}) that those jumps 
result in jumps of the atomic momentum while crossing 
a node of the SLW. If the value of the atomic energy is close to the separatrix 
one, the atom after the corresponding jump-like change in $p$ can either 
overcome the potential barrier and leave a potential well or  
it will be trapped by the well, or it will move as before. 
The jump-like behavior of $u$ is the ultimate reason of chaotic atomic 
walking along a rigid SW. 

The total atomic energy (\ref{H}) consists of the kinetic one, $K=\omega_rp^2/2$, 
and the potential one,  $U= -u\cos x -z\Delta/2$. The optical potential changes 
its depth in course of time. Averaging over fast oscillations of the 
internal atomic variables, we get the averaged potential
$\bar U= -\bar u\cos x -\bar z\Delta/2$ that can be used to explain    
why atoms move in such or another way. 

At small detunings $|\Delta| \ll 1$, the potential is approximately 
$U \simeq  u\cos x$. 
If $K(\tau=0)> |U_{\rm max}|=1$, then the atom will move 
ballistically. This occurs if the initial atomic 
momentum, $p_0$, satisfies to the condition $p_0> \sqrt{2/\omega_r}$. 
If the initial conditions are chosen to give 
$0\le K(\tau=0)+U(\tau=0) \le 1$, the corresponding atoms with 
$0 \lesssim p_0 \lesssim  \sqrt{2/\omega_r}$ are expected to move chaotically 
at the appropriate values of $\Delta$. 

\subsection{Focusing and splitting}
\pst

In this section we demonstrate how to focus and split atomic beams 
crossing a Gaussian laser beam by varying only one of the control
parameters, the atom-field detuning $\Delta$. 
Firstly, we perform simulation with a negligible probability 
of spontaneous emission and solve the Hamiltonian equations of motion (\ref{Hamsys}) 
at comparatively large value of the detuning, $\Delta=1$.
To be concrete we take as an example   
calcium atoms with the working intercombination transition $4^1S_0-4^3P_1$ 
at $\lambda_a = 657.5$~nm, the recoil frequency $\nu_{\rm rec}\simeq 10$~KHz, 
and the lifetime of the excited state $T_{\rm sp} = 0.4$~ms.
Taking the maximal Rabi frequency to be $\Omega_0/2\pi = 2\cdot 10^7$~Hz, 
the radius of the laser beam $r = 0.3$~cm, and the mean longitudinal velocity 
$v_z=10^3$~m/s, the interaction time is estimated to be $0.9$~ms, longer than 
the atomic lifetime. 
The normalized recoil frequency is $\omega_r = 4\pi \nu_{\rm rec}/
\Omega_0=10^{-3}$ and the normalized characteristic 
time is $\sigma_{\tau}=400$. 

\begin{figure}[!tpb]
\includegraphics[width=0.4\textwidth]{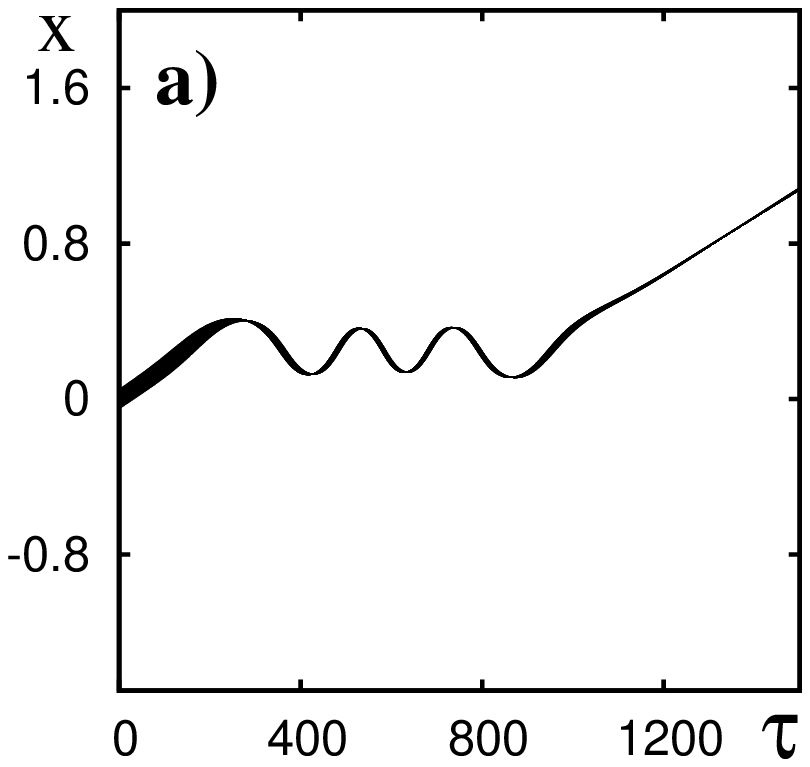}
\includegraphics[width=0.4\textwidth]{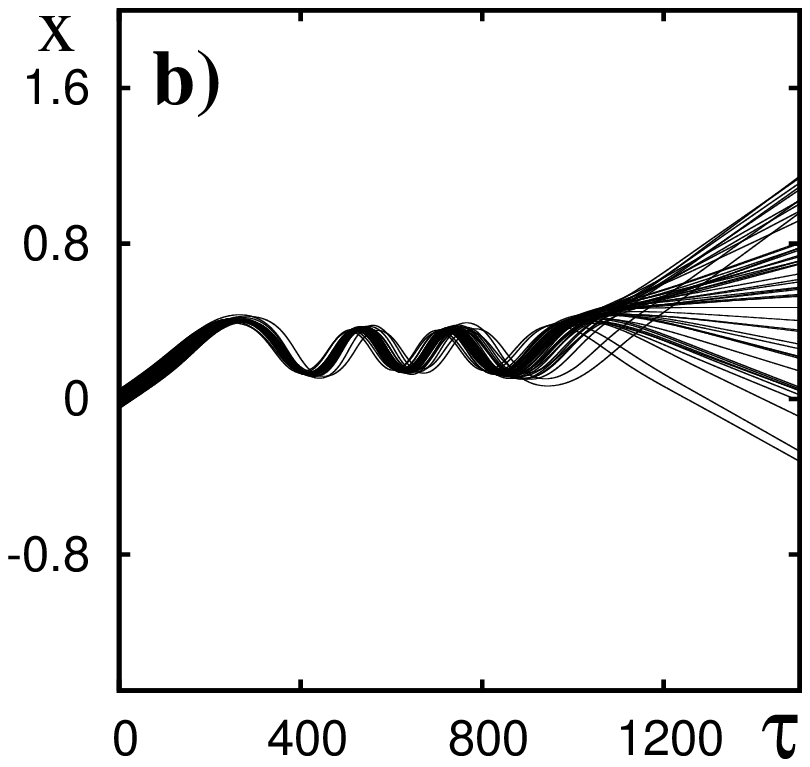}
\caption{(a) Focusing the atomic beam with a long lifetime of the excited state.
(b) The effect of spontaneous emission on the focusing. 
The detuning is $\Delta=1$ in both the cases. The 
atomic position $X$ is in units of the optical wavelength.}  
\label{fig1}
\end{figure}

Trajectories for 50 calcium atoms to be prepared in the ground states 
($u_0=v_0=0$, $z_0=-1$) with the same initial momentum,  
$p_0=10$, and initial positions in the range $-\pi/10\le x \le\pi/10$ 
are shown in Fig.~\ref{fig1}. In units of the optical wavelength, $X=x/2\pi$, 
this range is $-0.05\le X \le 0.05$. 
The focusing occurs at those moments of time 
when the average transverse momentum in the atomic beam is approximately equal 
to zero. If one turns off the laser at one of these moments, it becomes possible 
to reduce the beam width practically in ten times. The reason of focusing 
is simple. It is well known \cite{Letokhov} that at positive blue  
detunings atoms are attracted to the nodes of the SLW where the minima 
of the optical potential are situated at $\Delta>0$. The first node the 
atoms reach at $p_0>0$ is situated at $X=1/4$. The initial kinetic atomic 
energy, $K_0=0.05$, is not enough to overcome the potential barrier whose 
depth can be estimated to be $\simeq 0.35$ because the simulation gives 
$\bar u \simeq 0$ and $\bar z \simeq -0.7$. So, all 
the atoms in the beam oscillate in the first potential well in the $x$-direction 
around 
the first node. The initial width of the beam, $\delta X_0=0.1$, is gradually 
reduced because in course of time the atoms with initial negative positions catch up with the 
ones with positive $X_0$ near the first turning point 
where the average beam momentum is close to zero. The time interval of the atomic  
interaction with the SLW field 
is estimated to be $3\sigma_{\tau}=1200$. So, the atoms leave the potential well 
after that time and move freely (see Fig.~\ref{fig1}).

In order to take into account spontaneous emission 
we use the standard stochastic wave-function technique  
\cite{Carmichael,Dal,Dum} for solving Eqs.~(\ref{mainsys}). 
The integration time is divided into a large number of small time intervals 
$\delta\tau$. At the end of the first interval, $\tau=\tau_1$, the 
probability of spontaneous emission,  
$s_1=\gamma \delta \tau |a_{\tau_1}|^2/(|a_{\tau_1}|^2 + |b_{\tau_1}|^2)$, 
is computed and compared with a random number, $\varepsilon$, from the interval 
$[0,1]$. If $s_1 < \varepsilon_1$, then one prolongs the integration  
but renormalizes the state vector in the end of the first interval 
at $\tau= \tau_1^+$: $a_{\tau_1^+}=a_{\tau_1}/\sqrt{|a_{\tau_1}|^2 
+ |b_{\tau_1}|^2}$ and $b_{\tau_1^+}=b_{\tau_1}/\sqrt{|a_{\tau_1}|^2 + 
|b_{\tau_1}|^2}$. If $s_1 \ge \varepsilon_1$, then the atom emits a 
spontaneous photon and jumps to the ground state at $\tau= \tau_1$ 
with $A_{\tau_1}=\alpha_{\tau_1}=\beta_{\tau_1}=0$, $B_{\tau_1}=1$. Its momentum 
in the $x$ direction changes for a random number from the interval 
$[0,1]$ due to the photon recoil effect, and the next time step commences. 
  
We simulate lithium atoms with the relevant 
transition $2S_{1/2}-2P_{3/2}$, the corresponding wavelength
$\lambda_a = 670.7$~nm, recoil frequency $\nu_{\rm rec}=63$~KHz, 
and the decay time $T_{\rm sp} = 2.73 \cdot 10^{-8}$~s.
With the maximal Rabi frequency $\Omega_0/2\pi \simeq 126$~MHz 
and the radius of the laser beam $r = 0.05$~cm 
one gets $\omega_r = 10^{-3}$, $\sigma_{\tau}=400$, and $\gamma=0.05$. 
Trajectories for 50 spontaneously emitting atoms 
under the same conditions as in Fig.~\ref{fig1}a are shown in Fig.~\ref{fig1}b. 
As expected, spontaneous emission destroys in part the effect of focusing. However, 
the atoms move more or less coherently because spontaneous emission 
events are comparatively rare at $\Delta=1$. 
\begin{figure}[!tpb]
\includegraphics[width=0.4\textwidth]{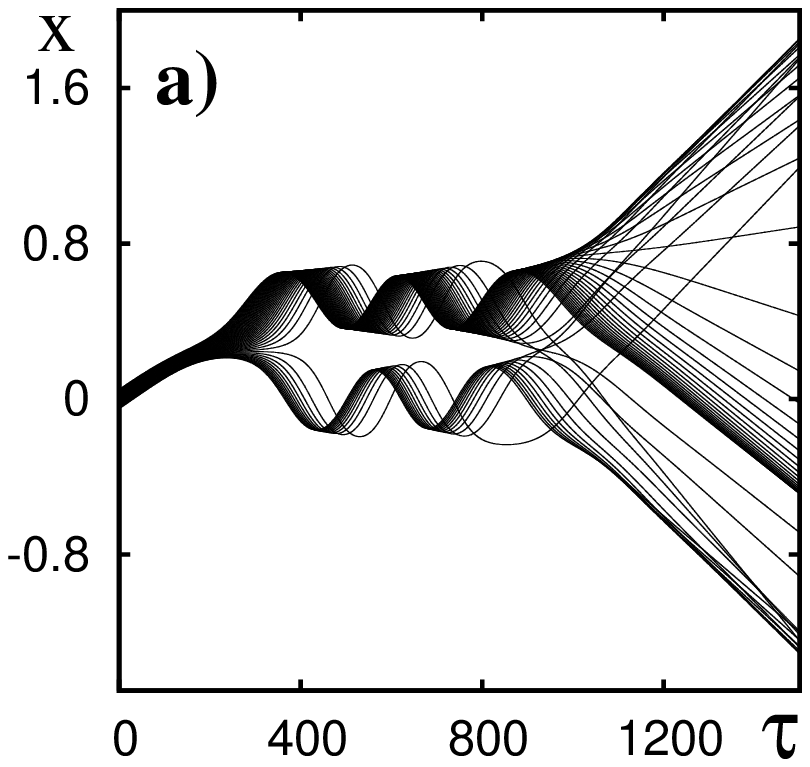}
\includegraphics[width=0.4\textwidth]{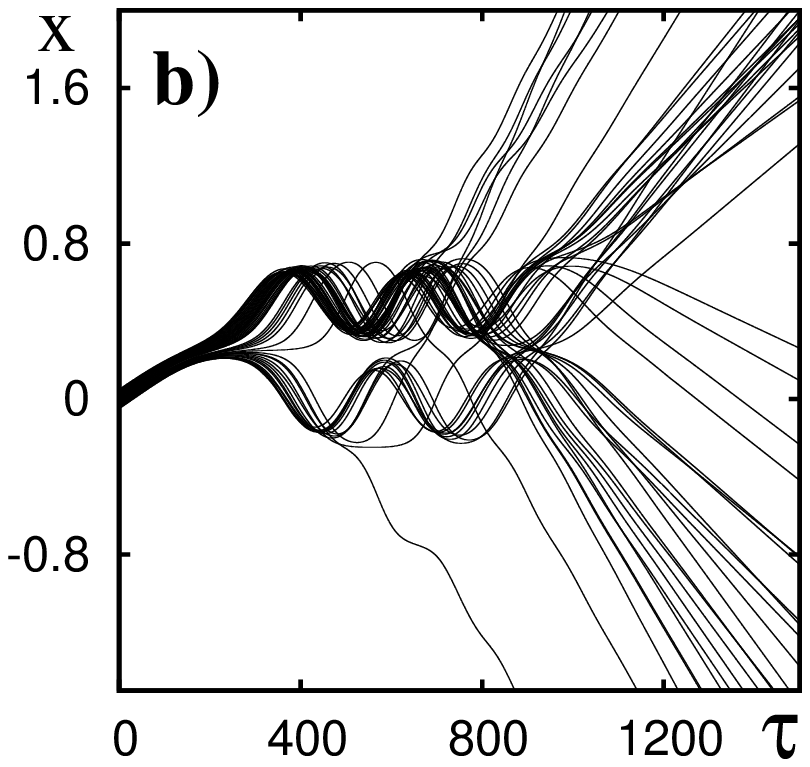}
\caption{Splitting the atomic beam (a) without and
(b) with spontaneous emission. The detuning is $\Delta=-1$.  }
\label{fig2}
\end{figure}              

The other effect, we would like to demonstrate with atomic beams crossing 
the SLW, is a splitting of the beam. To do this one needs to choose such the 
value of the detuning in order that some atoms in the beam would be trapped in the first well 
of the optical potential but another ones could overcome the barrier and leave 
that well. It is possible to split atomic beams as at positive and negative 
values of the detuning. As an example, we demonstrate in Fig.~\ref{fig2}   
the effect of splitting at $\Delta=-1$ for atoms without and with 
spontaneous emission. It is seen that spontaneous emission changes slightly 
the effect because a few atoms may leave the potential wells due to 
random recoils.

\section{Bunching and chaotic scattering of atoms}
\pst

\begin{figure}[!tpb]
\includegraphics[width=0.4\textwidth]{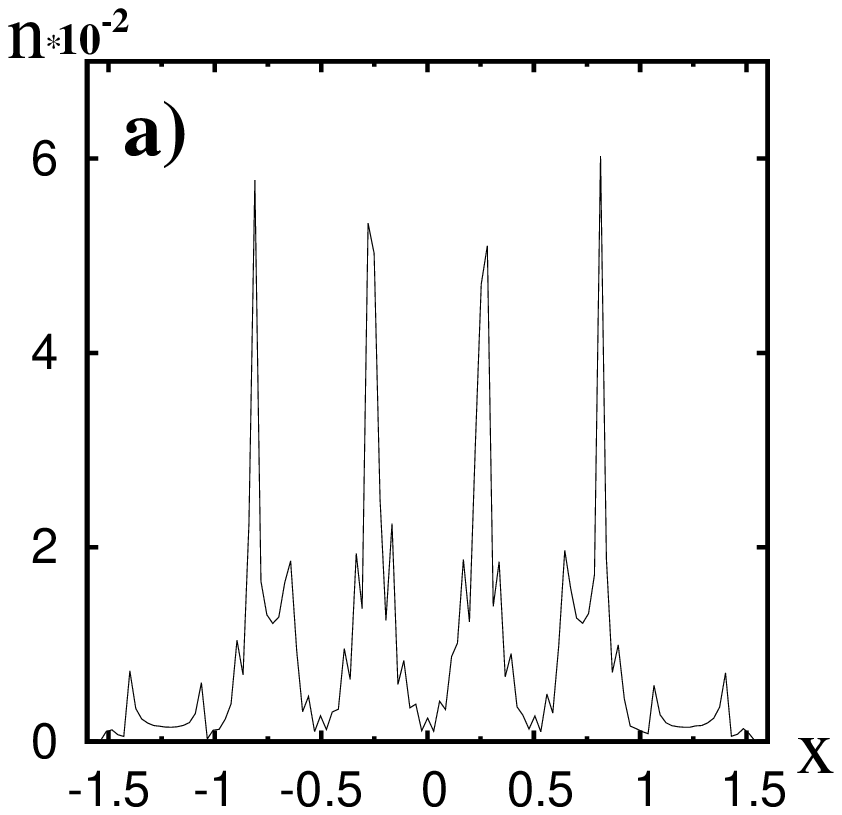}
\includegraphics[width=0.4\textwidth]{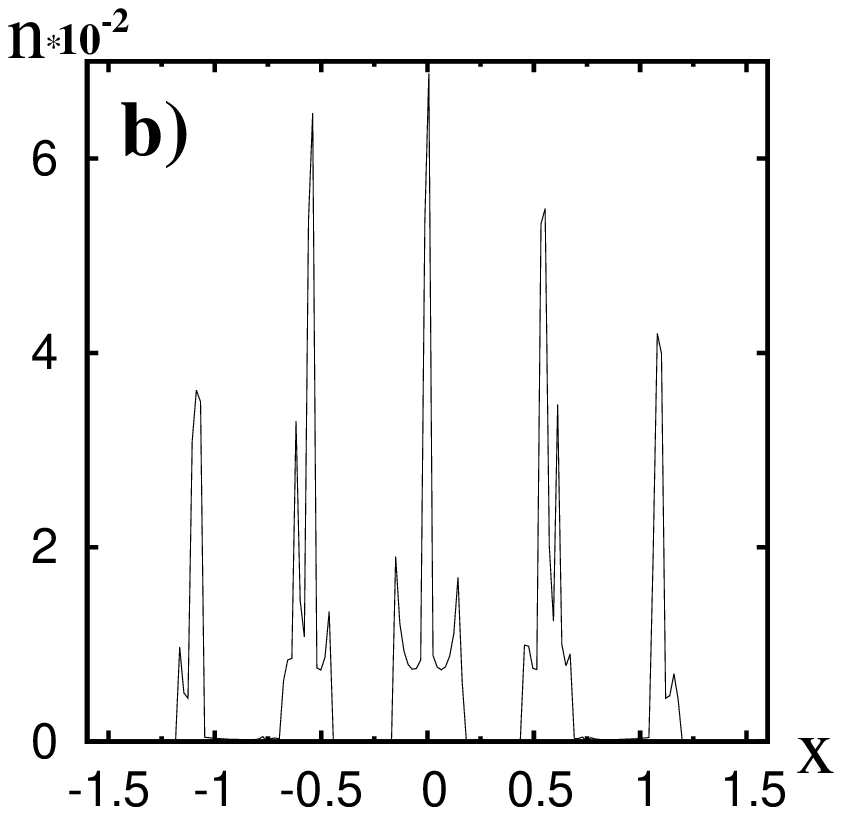}
\caption{The effect of bunching of $10^5$ calcium atoms around (a) 
the SLW nodes (blue detuning, $\Delta =1$) and (b) the antinodes (red detuning, 
$\Delta =-0.2$). The plot of atomic density $n=N(X)/N_0$ at the fixed moment of 
time $\tau=1000$.} 
\label{fig3}
\end{figure}

The ability of blue and red detuned lasers to attract atoms to the nodes 
and antinodes of the SLW, respectively, can be used to create periodic 
structures composed of atoms deposited on substrates in the process of 
optical nanolithography \cite{Sleator,Timp,McClelland}. 
To simulate a real experiment we consider a beam with  
$N_0=10^5$ calcium  atoms with the initial Gaussian distribution   
(with the rms $\sigma_x=\sigma_p=2$ and the average values $x_0=0$ and  
$p_0=10$) and compute their distribution against the SLW 
at a fixed moment of time $\tau=1000$. 
The bunching of atoms at the SLW nodes at $\Delta =1$ (blue detuning)
is shown in Fig.~\ref{fig3}a where the atomic density, $n=N(X)/N_0$, is plotted 
along the optical axis $X$ at $\tau=1000$. The same effect, but with the atoms 
bunching around the SLW antinodes (red detuning, $\Delta =-0.2$), is shown 
in Fig.~\ref{fig3}b. In both the cases we get a periodic atomic relief with 
the period $\lambda_f/2$ the width of which is restricted by the time the atoms 
interact with the Gaussian laser beam. 

The problem we consider resembles   
the scattering process with particles entering an interaction region along 
completely regular trajectories and leaving it along asymptotically 
regular trajectories. It is known from many studies in celestial mechanics,  
fluid dynamics and other disciplines that under 
certain conditions the motion inside the 
interaction region may have features that are typical for dynamical chaos, 
(homoclinic and heteroclinic tangles, fractals, strange invariant sets, 
positive finite-time Lyapunov exponents, etc.) although the particle's  
trajectories are not chaotic in a rigorous sense because chaos is strictly defined as 
an irregular motion over infinite time. It has been found \cite{Gaspard,
M03,PhysD,JETP04} that transient Hamiltonian chaos 
in the interaction region occurs due to existence of, at least, one nonattractive 
chaotic invariant set consisting of an infinite number of localized unstable 
periodic orbits and aperiodic orbits. This set possesses stable and unstable 
manifolds extending in the phase space into the regions of regular motion. The particles with the 
initial positions close to the stable manifold follow the chaotic-set 
trajectories for a comparatively long time, then deviate 
from them, and leave the interaction region along the unstable manifold. 
It is the common mechanism of chaotic scattering    
that in our problem causes the chaotic walking of atoms along the SLW. 
\begin{figure}[!tpb]
\includegraphics[width=0.4\textwidth]{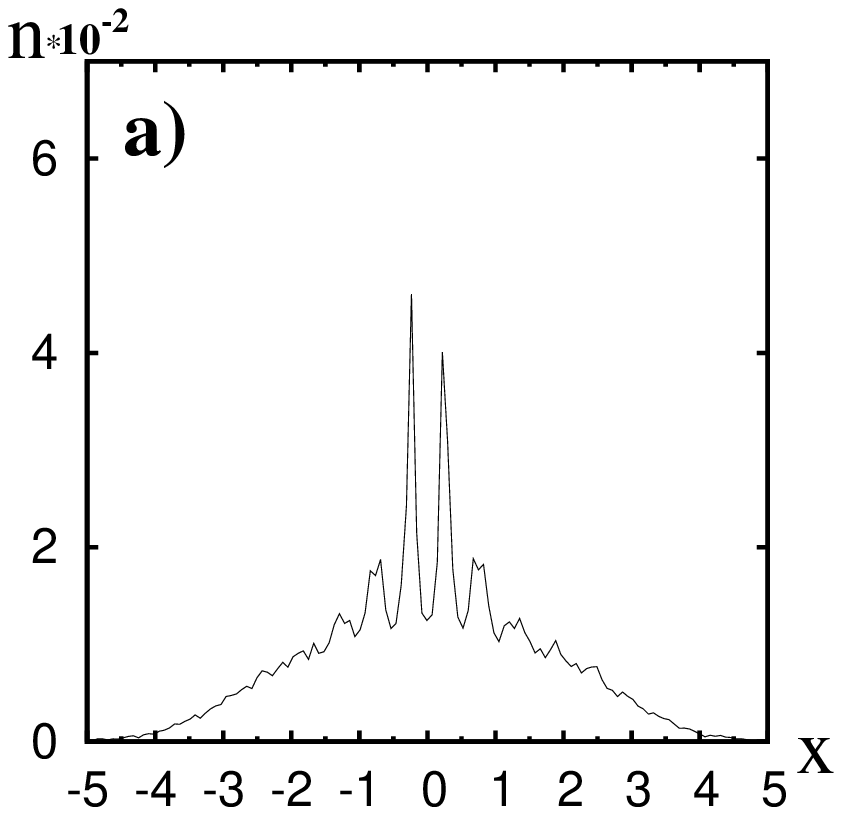}
\includegraphics[width=0.4\textwidth]{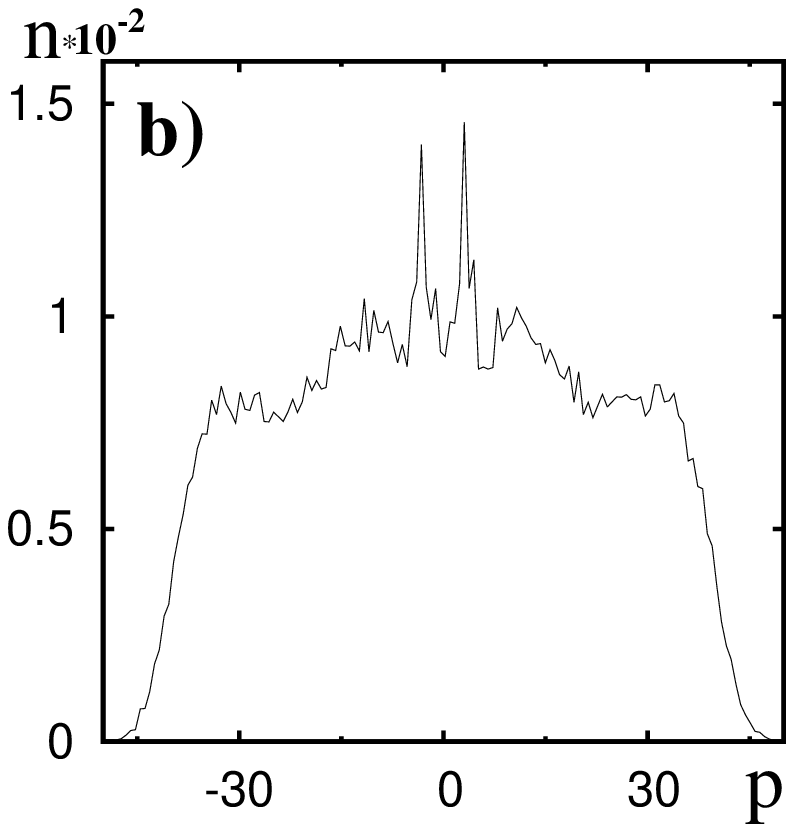}
\caption{The distributions of $10^5$ calcium atoms at $\tau=1000$ 
in (a) the real and (b) momentum space under the conditions 
of chaotic scattering at $\Delta=0.2$.}  
\label{fig4}
\end{figure}              
\begin{figure}[!tpb]
\includegraphics[width=0.4\textwidth]{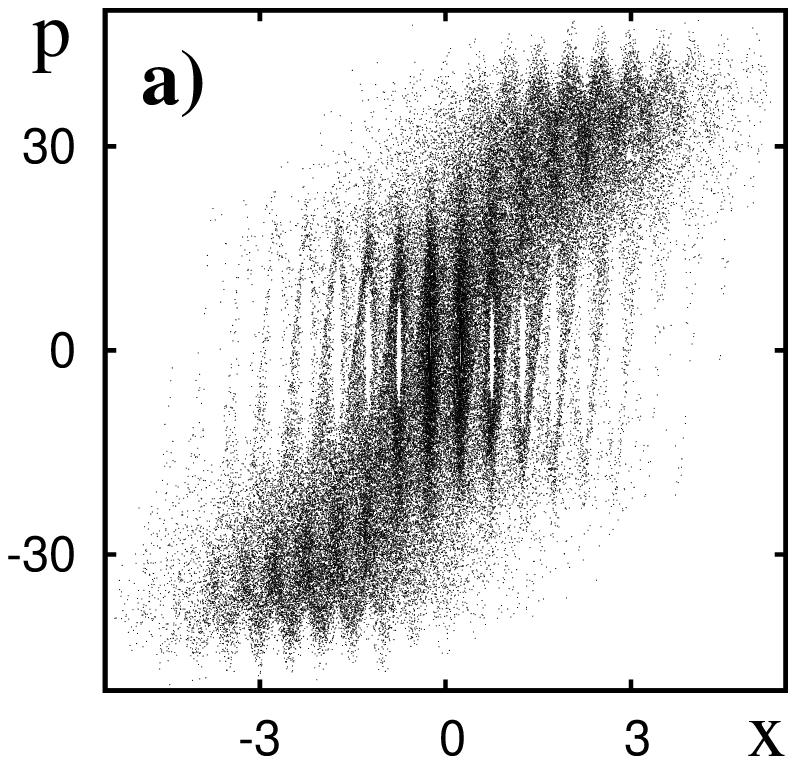}
\includegraphics[width=0.4\textwidth]{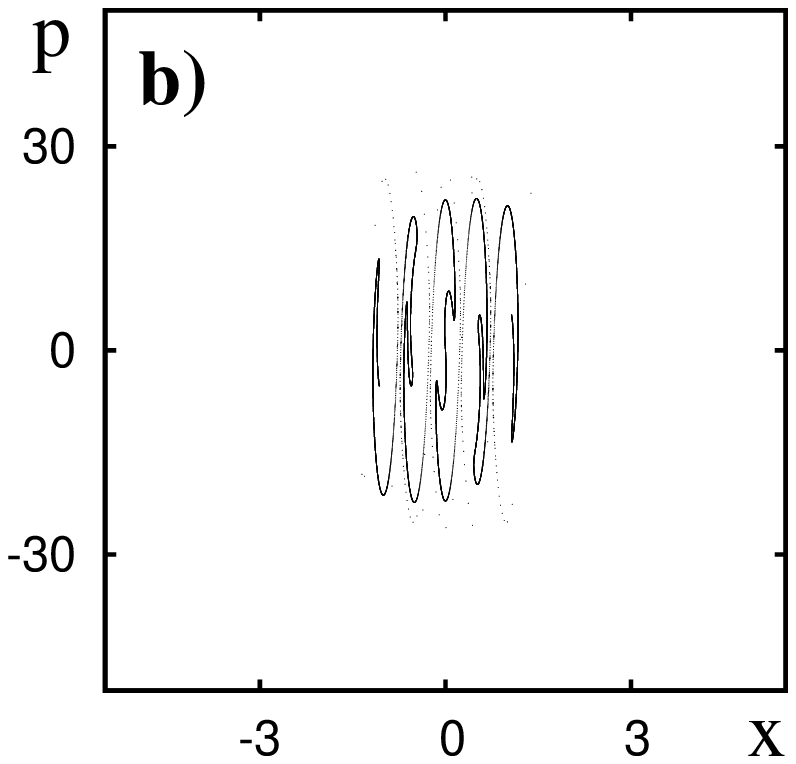}
\caption{Comparison of the distributions of $10^5$ calcium atoms 
at $\tau=1000$ over the phase plane in the regimes of  
(a) chaotic ($\Delta=0.2$) and (b) regular scattering ($\Delta=-0.2$).}  
\label{fig5}
\end{figure}              

In Fig.~\ref{fig4}a we show the atomic position distribution    
at $\tau=1000$ in the regime of the chaotic scattering at $\Delta =0.2$
with $10^5$ calcium  atoms. This plot should be compared with 
Fig.~\ref{fig3}a where the atomic position distribution is shown    
for regularly scattered atoms at $\Delta =1$. First of all, the distribution 
of chaotically scattered atoms has a prominent pedestal and is much broader. 
Moreover, it has no such a periodic structure as shown in Fig.~\ref{fig3}a. 
Only the peaks around the first two SLW nodes are prominent in Fig.~\ref{fig4}a.
The atomic position distribution in the momentum space in Fig.~\ref{fig4}b is much broader 
than the one for regularly scattered atoms at $\Delta =1$ (not shown).
Thus, we predict that under the conditions of chaotic scattering  
there should appear less contrast and more broadened atomic reliefs as 
compared to the case of regular 
scattering because a large number of atoms are expected to be deposited 
between the nodes as a result of chaotic walking along the SLW axis. 
The effect is expected to be more prominent under the coherent evolution  
but it seems to be observable with spontaneously emitting atoms as well.  
The difference between chaotic and regular scattering of atoms at a 
rigid SLW is especially prominent on the corresponding 
phase space portraits shown in Fig.~\ref{fig5} where positions and momenta of 
$10^5$ calcium  atoms are plotted at the fixed time moment.

\section{Conclusion}
\pst

We have simulated some geometric optics effects with atomic beams crossing  
a SLW in the limit of long relaxation time and with spontaneous emission 
taken into account.  Trajectories of spontaneously emitting atoms 
have been simulated with the help of the standard stochastic wave-function technique  
\cite{Carmichael,Dal,Dum}. It has been shown that by adjusting the detuning it is possible 
to focus, split and scatter atoms. The effects have been explained 
by a coupling between external and internal atomic degrees of freedom. 
The depth of the optical potential depends on the sign and value of 
the detuning. Varying $\Delta$, one can create conditions for focusing, 
splitting and bunching the atoms. It is remarkable that
near the atom-field resonance we have found the new type 
of atomic diffraction at a SLW without any modulation of its parameters  
that can be observed in real experiments. That would 
be the prove of existence of the novel 
type of atomic motion, chaotic walking in a deterministic environment. 
The effects found could be used in optical nanolithography to fabricate complex 
atomic structures on substrates. 

We predict that experiments on the scattering of atomic beams at a SLW  
can directly image chaotic walking of atoms along the SLW. 
In a real 
experiment the final spatial distribution can be recorded via fluorescence or 
absorption imaging on a CCD, commonly used methods in atom optics experiments 
yielding information on the number of atoms and the cloud's spatial size. 
The other possibility is a nanofabrication 
where the atoms after the interaction with the SW are deposited on a silicon 
substrate in a high vacuum chamber. In this case the spatial distribution 
can be analyzed with an atomic force microscope. As to the momentum 
distribution, it can be measured, for example, by a time-of-flight 
technique \cite{Raizen}. The modern tools of atom optics enable to create 
narrow initial atomic distributions in position and momentum, reduce coupling 
to the environment and technical noise, create one-dimensional optical 
potentials, and to measure spatial and momentum distributions 
with high sensitivity and accuracy.

\section*{Acknowledgments} 
This work was supported  by the Russian Foundation for Basic Research
(projects nos. 09-02-00358 and 09-02-01258), by the Integration grant from the Far-Eastern 
and Siberian branches of the Russian Academy of Sciences, and by the Program
``Fundamental Problems of  Nonlinear Dynamics''.

\end{document}